\documentclass[preprint,showpacs,preprintnumbers,amsmath,amssymb]{revtex4}

\usepackage{graphicx}
\usepackage{dcolumn}
\usepackage{bm}

\begin{document}


\title{Stability of Proton and Maximally Symmetric Minimal Unification Model \\
for Basic Forces and Building Blocks of Matter }
\author{Yue-Liang Wu }
 \affiliation{Institute of Theoretical Physics, Chinese Academy of sciences,
 Beijing 100080, China  }
 %
\date{$\ $ e-mail: ylwu@itp.ac.cn}
\begin{abstract}
With the hypothesis that all independent degrees of freedom of
basic building blocks should be treated equally on the same
footing and correlated by a possible maximal symmetry, we arrive
at an 4-dimensional space-time unification model. In this model
the basic building blocks are Majorana fermions in the spinor
representation of 14-dimensional quantum space-time with a gauge
symmetry $G_M^{4D} = SO(1,3)\times SU(32)\times U(1)_A \times
SU(3)_F$. The model leads to new physics including mirror
particles of the standard model. It enables us to issue some
fundamental questions that include: why our living space-time is
4-dimensional, why parity is not conserved in our world, how is
the stability of proton, what is the origin of CP violation and
what can be the dark matter.
\end{abstract}

\maketitle

The most important issues in elementary particle physics concern
fundamental questions such as: what is the basic building blocks of
nature? what is the basic symmetries of nature? what is the basic
forces of nature? why our living space-time is 4-dimensional? why
parity is not conserved in our world? how is the stability of
proton? what can be the dark matter? what is origin of CP violation?
why neutrinos are so light? In the standard model\cite{SM1,SM2,SM3},
quarks\cite{GM,GZ} and leptons are regarded as the basic building
blocks and described by the gauge symmetries $U(1)_Y \times SU(2)_L
\times SU(3)_C$ and Lorentz symmetry SO(1,3). The gauge bosons
corresponding to the symmetries mediate interactions among quarks
and leptons via electromagnetic, weak and strong forces. The local
Lorentz symmetry SO(1,3) reflects gravitational force.

The standard model has been tested by more and more precise
experiments at the energy scale of order 100 GeV. The weak
interaction is well described by the left-handed $SU(2)_L$ gauge
symmetry based on the fact of parity (P) noninvariance\cite{PV}. The
strong interaction described by the Yang-Mills gauge theory\cite{YM}
with symmetry $SU(3)_C$ displays a behavior of asymptotic
freedom\cite{AF1,AF2}. The standard model has been shown to be a
renormalizable quantum field theory\cite{DR}. The strength of all
forces has been turned out to run into the same magnitude at high
energy scales\cite{RC}, which makes it more attractive for the
exploration of grand unification
theories\cite{SU422,SU5,SO10G,SO10}. SU(5) gauge model\cite{SU5} is
known to be a minimal grand unification theory. For massive
neutrinos, SO(10) model\cite{SO10G,SO10} may be regarded to be a
minimal one. An interesting feature in the SO(10) model is that all
the quarks and leptons in each family can be unified into a single
spinor representation. One of the important predictions in either
SU(5) or SO(10) model is proton decay. Namely proton is no longer a
stable particle in the SU(5) and SO(10) models. As a consequence,
the minimal SU(5) and SO(10) models have been strongly constrained
by the current experimental data on proton decays.

In here we shall explore other possible unification models. For
that, let us reanalyze what symmetry means in the grand
unification models. We shall first examine the symmetries in the
standard model. It is seen that symmetries are introduced to
establish the relations among different quantum charges of quarks
and leptons. The known quantum charges of quarks and leptons
consist of isospin charges, color charges, lepton charges, spin
charges and chiral-boost charges. Specifically, $SU(2)_L$ is
introduced to describe the symmetry between two isospin charges,
$SU(3)_C$ characterizes the symmetry among three color charges,
SO(1,3) reflects the symmetry among 2 spin charges and 2
chiral-boost charges. In the SO(10) model, SO(10) characterizes
the symmetry of unified isospin-color-lepton charges. When
treating all the quantum charges on the same footing\cite{SO14},
we arrive at a symmetry group $SO(1,3)\times SO(10)$, which may be
regarded as a symmetry of 14-dimensional (14D) {\it quantum
space-time}. This is because its dimensions are determined by the
basic quantum charges of quarks and leptons (2 spin and 2 boost
charges, 2 isospin charges, 3 color and 3 anticolor charges,
lepton and antilepton charges). The independent degrees of freedom
of quarks and leptons are given by the spinor representation of
14D quantum space-time. As there are 64 real independent degrees
of freedom for each family quarks and leptons, the symmetry only
based on the quantum charges of quarks and leptons cannot be a
maximal symmetry that establishes possible correlations among the
independent degrees of freedom of quarks and leptons. In other
words, a large amount of independent degrees of freedom of quarks
and leptons are not related via the symmetry group $SO(1,3)\times
SO(10)$. It then becomes manifest that in the grand unification
models one only considers symmetries among the basic quantum
charges of building blocks rather than among all independent
degrees of freedom of building blocks. In the present paper, we
shall extend the usual grand unification models by considering a
possible maximal symmetry among all independent degrees of freedom
of basic building blocks.

As a principle in our present consideration, we make a simple
hypothesis that {\it all independent degrees of freedom of basic
building blocks should be treated equally on the same footing and
correlated by a possible maximal symmetry in a minimal unified
scheme.} For convenience of mention, we may refer such a
hypothesis as a maximally symmetric minimal unification hypothesis
(MSMU-hypothesis), and the resulting model as a maximally
symmetric minimal unification model (MSMUM).

In order to establish possible correlations among all independent
degrees of freedom of basic building blocks, we can infer from the
above MSMU-hypothesis the following deduction: {\it fermions as
basic building blocks of nature should be Majorana fermions in the
spinor representation of high dimensional quantum space-time which
is determined by the basic quantum charges of building blocks. The
chirality of basic building blocks should be well-defined when
parity is considered to be a good symmetry}. This deduction
implies that the dimensions of quantum space-time should allow a
spinor representation for both Majorana and Weyl fermions, which
requires the quantum space-time to be at the dimensions D = 2 + 4n
$(n=1,2, \cdots )$, i.e., D = 2, 6, 10, 14, 18, 22, 26, $\cdots$.

  In the spirit of MSMU-hypothesis, the minimal dimension
needed for a MSMUM is D=14. Thus the basic building blocks are
Majorana fermions in the spinor representation of 14D quantum
space-time. Namely each family of Majorana fermion has $128=2^{7}$
independent real degrees of freedom, which is twice to the quarks
and leptons in the standard model. With the Majorana condition in
the spinor representation of 14D quantum space-time, we will
arrive at an interesting 4D space-time MSMUM with a gauge symmetry
for each family $G_M^{4D} = SO(1,3)\times SU(32)\times U(1)_A$.

For an explicit demonstration, let us denote $\Psi$ as fermionic
building block in the spinor representation of 14D space-time. The
Majorana condition implies that
\begin{equation}
\Psi = \Psi^{\hat{c}} = \hat{C} \bar{\Psi}^{T}
\end{equation}
Here $\Psi^{\hat{c}}$ defines the charge conjugation in the 14D
quantum space-time. The 128-dimensional spinor representation of
Majorana fermion $\Psi$ is found to have the form
\begin{equation}
\Psi = \left( \begin{array}{c} F_{L} + F'_R \\ F_{R} + F'_L
\end{array} \right)
\end{equation}
with $F_{L,R}$ defined as
\begin{eqnarray}
F_{L,R}^{T} & & = [\ U_{r}, U_{b}, U_{g}, N, D_{r}, D_{b}, D_{g},
E,  \nonumber \\
& & D_{r}^{c}, D_{b}^{c}, D_{g}^{c}, E^{c}, -U_{r}^{c}, -U_{b}^{c},
-U_{g}^{c}, -N^{c}\  ]_{L,R}^{T} \nonumber \\
F_{L,R}^{'T} & & =  [\ U'_{r}, U'_{b}, U'_{g}, N', D'_{r}, D'_{b},
D'_{g}, E',  \\
& & D_{r}^{'c}, D_{b}^{'c}, D_{g}^{'c}, E^{'c}, -U_{r}^{'c},
-U_{b}^{'c}, -U_{g}^{'c}, -N^{'c} \ ]_{L,R}^{T} \nonumber
\end{eqnarray}
and the charge conjugation matrix is given by
\begin{eqnarray}
& & \hat{C} = i \sigma_{1}\otimes \sigma_{2}\otimes \sigma_{2}
\otimes 1\otimes 1 \otimes \sigma_{3} \otimes \sigma_{2}
\end{eqnarray}
which satisfies $\hat{C}^{\dagger}  =  \hat{C}^{-1} = -\hat{C}^{T}
= -\hat{C}$ and $\hat{C}\hat{C}^{\dagger}=1$.  All the fermions
$\psi = U_{i},\ D_{i},\ E,\ N, \cdots $ are four complex component
fermions defined in the 4D. The indices ``L'' and ``R'' denote the
left-handed and right-handed fermions in 4D, i.e., $\psi_{L,R} =
\frac{1}{2}(1\mp \gamma_{5}) \psi$. The index ``c'' represents the
charge conjugation in 4D, $\psi^{c} = C\bar{\psi}^{T}$ with
$C=i\gamma_{0}\gamma_{2}=i\sigma_{3}\otimes \sigma_{2}$.

For convenience of discussions, we present the explicit form of
the gamma matrices $\hat{\Gamma}_{\hat{I}}=(\gamma_{a},
\Gamma_{I})$ in the spinor representation of 14D quantum
space-time
\begin{eqnarray}
\gamma_{0} & = & 1\otimes 1\otimes 1\otimes 1\otimes 1
\otimes \sigma_{1}\otimes 1\ , \nonumber \\
\gamma_{1} & = & i\ 1\otimes 1\otimes 1\otimes 1\otimes 1
\otimes \sigma_{2}\otimes \sigma_{1}\ , \nonumber \\
\gamma_{2} & = & i\ 1\otimes 1\otimes 1\otimes 1\otimes 1
\otimes \sigma_{2}\otimes \sigma_{2}\ , \nonumber \\
\gamma_{3} & = & i\ 1\otimes 1\otimes 1\otimes 1\otimes 1
\otimes \sigma_{2}\otimes \sigma_{3}\ , \nonumber \\
\Gamma_{1} & = & \ \sigma_{1}\otimes \sigma_{1}\otimes 1
\otimes 1\otimes \sigma_{2}\otimes \sigma_{3}\otimes 1\ , \nonumber \\
\Gamma_{2} & = & \ \sigma_{1}\otimes \sigma_{2}\otimes 1
\otimes \sigma_{3}\otimes \sigma_{2}\otimes \sigma_{3}\otimes 1\ , \nonumber \\
\Gamma_{3} & = & \ \sigma_{1}\otimes \sigma_{1}\otimes 1
\otimes \sigma_{2}\otimes \sigma_{3}\otimes \sigma_{3}\otimes 1\ , \\
\Gamma_{4} & = & \ \sigma_{1}\otimes \sigma_{2}\otimes 1
\otimes \sigma_{2}\otimes 1\otimes \sigma_{3}\otimes 1\ , \nonumber \\
\Gamma_{5} & = & \ \sigma_{1}\otimes \sigma_{1}\otimes 1
\otimes \sigma_{2}\otimes \sigma_{1}\otimes \sigma_{3}\otimes 1\ , \nonumber \\
\Gamma_{6} & = & \ \sigma_{1}\otimes \sigma_{2}\otimes 1
\otimes \sigma_{1}\otimes \sigma_{2} \otimes \sigma_{3}\otimes 1\ , \nonumber \\
\Gamma_{7} & = & \ \sigma_{1}\otimes \sigma_{3}\otimes \sigma_{1}
\otimes 1\otimes 1\otimes \sigma_{3}\otimes 1\ , \nonumber \\
\Gamma_{8} & = & \ \sigma_{1}\otimes \sigma_{3}\otimes \sigma_{2}
\otimes 1\otimes 1 \otimes \sigma_{3}\otimes 1\ , \nonumber \\
\Gamma_{9} & = & \ \sigma_{1}\otimes \sigma_{3}\otimes \sigma_{3}
\otimes 1\otimes 1\otimes \sigma_{3}\otimes 1\ , \nonumber \\
\Gamma_{10} & = & \ \sigma_{2}\otimes 1\otimes 1\otimes 1\otimes
1\otimes \sigma_{3}\otimes 1\  \nonumber
\end{eqnarray}
where $\sigma_{i}$ ($i=1,2,3$) are Pauli matrices, and ``1'' is
understood as a $2\times 2$ unit matrix. We may define two gamma
matrices from $\hat{\Gamma}_{\hat{I}}$ ($\hat{I}=0$, $\cdots$, 13)
as
\begin{eqnarray}
\Gamma_{15} & = &  \hat{\Gamma}_{0} \cdots \hat{\Gamma}_{13} =
\sigma_{3}\otimes 1\otimes 1\otimes 1\otimes 1 \otimes \sigma_3 \otimes 1  \nonumber \\
\Gamma_{11} & = & \Gamma_{1} \cdots \Gamma_{10} =
\sigma_{3}\otimes 1\otimes 1\otimes 1\otimes 1 \otimes 1\otimes 1
\end{eqnarray}
with $\Gamma_{15}\hat{\Gamma}_{\hat{I}} = -
\hat{\Gamma}_{\hat{I}}\Gamma_{15}$ and $\Gamma_{11}\Gamma_{I} = -
\Gamma_{I}\Gamma_{11}$. The Weyl representation of $\Psi$ is
defined through the projector operators
$P_{W,E}=(1\mp\Gamma_{15})/2$ with $P_{W,E}^{2} = P_{W,E}$
\begin{eqnarray}
\Psi_{W} & = & P_{W}\Psi = \frac{1}{2}(1 - \Gamma_{15})\Psi \equiv
\left( \begin{array}{c} F_{L} \\ F_{R}
\end{array} \right),  \nonumber \\
\Psi_{E} & = & P_{E}\Psi = \frac{1}{2}(1 + \Gamma_{15})\Psi \equiv
 \left( \begin{array}{c} F'_{R}
\\ F'_{L}
\end{array} \right)
\end{eqnarray}
Here $\Psi_W$ and $\Psi_E$ may be mentioned as 'westward' and
'eastward' fermions in order to distinguish with the left- and
right- handed fermions defined via the projector operators
$P_{L,R} = (1 \mp \gamma_{5})/2$ in 4D space-time.

The westward fermion $\Psi_{W}$ is a Majorana-Weyl fermion, it
contains 64 independent degrees of freedom that exactly represent
quarks and leptons in each family with right-handed neutrino. The
eastward Majorana-Weyl fermion $\Psi_{E}$ may be regarded as
mirror particles and mentioned as mirroquarks and mirroleptons.

We can construct following tensors by $\Gamma$-matrices
\begin{eqnarray}
& & T^{U} \equiv ( \Sigma^{IJ}, \  \Gamma_{11}
\Sigma^{IJKL},\ \Gamma_{11} ) \nonumber  \\
& & T_5^{V} \equiv (\Sigma^{IJK},\  i\Gamma_{11} \Sigma^{IJK}),\
 T^{\hat{V}} \equiv (T^{U}, \ i\Gamma_{15} T_5^{V})  \\
& & T_5^{W} \equiv (\Sigma^{I},\ i\Gamma_{11} \Sigma^{I},\
\Sigma^{IJKLM} ), \
 T_5^{\hat{W}} \equiv ( T_5^{W}, \Gamma_{15} T^{U}) \nonumber
\end{eqnarray}
with $\Sigma^I \equiv \frac{1}{2} \Gamma^I$, $\Sigma^{IJ} =
\frac{i}{4} [ \Gamma^I , \Gamma^J ]$ and others are the higher
order antisymmetric tensors. Under charge conjugation and parity
transformation, we have
\begin{eqnarray}
 & & \hat{C} T^{\hat{V}} \hat{C}^{\dagger}= - (T^{\hat{V}})
^{T}, \quad  \Gamma^0 T^{\hat{V}}
 \Gamma^0 = (T^{\hat{V}})^{\dagger}  \nonumber \\
& & \hat{C} T_5^{\hat{W}} \hat{C}^{\dagger}= (T_5^{\hat{W}}) ^{T},
\quad \Gamma^0 T_5^{\hat{W}}
 \Gamma^0 = -(T_5^{\hat{W}})^{\dagger} \\
 && \hat{C} \Sigma^{ab} \hat{C}^{\dagger}= - (\Sigma^{ab}) ^{T},
\quad \Gamma^0 \Sigma^{ab} \Gamma^0 = (\Sigma^{ab})^{\dagger}
\nonumber
\end{eqnarray}
where $\Sigma^{ab} = \frac{1}{4i} [ \gamma^a , \gamma^a ] $ ($a, b
= 0,\ 1,\ 2,\ 3$) are the generators of Lorentz group SO(1,3).
Here the matrices
\begin{eqnarray}
T^{\hat{A}} = (T^{\hat{V}}, T_5^{\hat{W}})\ , \quad ( \hat{A} = 1,
\cdots , 1024 )
\end{eqnarray}
form the generators of symmetry group $SU(32)\times U(1)_A$. Where
$T^{\hat{V}}$ $( \hat{V} = 1, \cdots , 496)$ form the generators
of subgroup $SO(32)$, $T^{U}$ ($U = 1, \cdots , 256$) the
generators of a subgroup $SU(16)\times U(1)$ and $\Sigma^{IJ}$ the
generators of a subgroup SO(10).

It is of interest to observe that although the 128-dimensional
spinor representation of Majorana fermion $\Psi$ is defined in the
14D quantum space-time, while its motion cannot realize in the
whole 14D space-time corresponding to the 14D quantum space-time.
This is because no kinetic term can exist in the corresponding 10D
space, and its motion can only be emergent in an 4D space-time out
of the 14D space-time. This can be seen from the identities
\begin{eqnarray}
& & \bar{\Psi} \Gamma^{I}i\partial_{I} \Psi \equiv \frac{1}{2}
 \partial_{I} \left( \bar{\Psi} i \Gamma^{I} \Psi \right)  \nonumber \\
& & \bar{\Psi} \gamma^{a}i\partial_{a} \Psi \equiv \frac{1}{2} [
 \bar{\Psi} \gamma^{a}i\partial_{a} \Psi - i\partial_{a} (\bar{\Psi}) \gamma^{a}
 \Psi ]
\end{eqnarray}
with $a=0,1,2,3$, and $I = 1, \cdots , 10$. Here the Majorana
condition has been used. As there exists no motion in the 10D
space, the 32-dimensional spinor representation of 10D quantum
space is found to have a maximal symmetry $SU(32)\times U(1)_A$ in
the kinetic 4D space-time.

The number of families remains a puzzle. When treating the
observed three families equally on the same footing, a maximal
family symmetry among them is $SU(3)_F$ with the chiral-valued
generators
\begin{eqnarray}
T^{u} = ( \lambda^{i}, \  \gamma_5 \lambda^{s} ), \quad
(\lambda^i)^T = - \lambda^i, \ \  (\lambda^s)^T = \lambda^s
\end{eqnarray}
and $(i=1,2,3; \ s= 1, \cdots , 5)$. Where $\lambda^i$ form the
generators of $SO(3)_F \in SU(3)_F$.

We are now in the position to write down the Lagrangian of MSMUM
in the kinetic 4D space-time for the fermionic building blocks
\begin{eqnarray}
{\cal L}^{4D}_{F} & = & \frac{1}{2}\bar{\Psi} \gamma^{a}i D_{a}
\Psi \equiv \frac{1}{2}\bar{\Psi} \hat{\gamma}^{a}E^{\mu}_a i
D_{\mu} \Psi
\end{eqnarray}
which is self-Hermician. Here $E^{\mu}_a$ is introduced as frame
fields. The covariant derivative $D_{\mu}$ is defined as
\begin{eqnarray}
& & D_{\mu} = \partial_{\mu} - i \omega_{\mu}^{ab} \Sigma_{ab} - i
g_U \Omega_{\mu}^{\hat{A}} T^{\hat{A}} - i g_F F_{\mu}^{u} T^{u}
\end{eqnarray}
The above Lagrangian has a maximal gauge symmetry
\begin{eqnarray}
G_{M}^{4D} = SO(1,3)\times SU(32)\times U(1)_A \times SU(3)_F
\end{eqnarray}
where the three families $\Psi = (\Psi_1 , \ \Psi_2 ,\ \Psi_3 )$
belong to the fundamental representation of $SU(3)_F$.

A minimal scalar field that couples to fermions is in the
representations $\Phi_F = \gamma_0 \lambda^s \Phi_s^{\hat{W}}
T_5^{\hat{W}} \equiv \gamma_0 \lambda^s ( \gamma_5
\Sigma_s^{\hat{W}} + i \Pi_s^{\hat{W}} ) T_5^{\hat{W}}$ ($s=0,1,
\cdots , 5$; $\hat{W} = 1, \cdots , 527$). The self-Hermician
Lagrangian for Yukawa interactions is
\begin{eqnarray}
{\cal L}^{4D}_{Y} & = & \frac{1}{2}g_Y \Psi^{\dagger} \Phi_F \Psi
   =  \frac{1}{2}g_Y \bar{\Psi}  \lambda^s \Phi_s^{\hat{W}} T_5^{\hat{W}} \Psi
\end{eqnarray}
Here the minimal scalar field $\Phi_F$ contains representations
$(272)_{W,E} = (10 + \bar{10} + 126 + \bar{126})_{W,E}$ that are
needed for generating masses of quarks and leptons as well as
their mirroparticles via spontaneous symmetry breaking.

A scalar field that interacts with gauge bosons can be different
from the one that couples to fermions. Such that a scalar field in
the adjoint representation of SU(32) $\hat{\Phi} \equiv
\hat{\Phi}^{\hat{A}} T^{\hat{A}}$ only couples to gauge bosons. The
self-Hermician Lagrangian for minimal scalar fields is given by
\begin{eqnarray}
{\cal L}^{4D}_{S} & = & Tr D_{\mu}\Phi_F^{\dagger}D_{\mu}\Phi_F +
Tr D_{\mu}\hat{\Phi}^{\dagger}D_{\mu}\hat{\Phi}
\end{eqnarray}

In general, a total self-Hermician Lagrangian is
\begin{eqnarray}
{\cal L}^{4D} = {\cal L}^{4D}_{F} + {\cal L}^{4D}_{G} + {\cal
L}^{4D}_{S} + {\cal L}^{4D}_{Y} + {\cal L}^{4D}_{H}
\end{eqnarray}
Here ${\cal L}^{4D}_{G}$ represents the Yang-Mills gauge
interactions and ${\cal L}^{4D}_{H}(\hat{\Phi},\Phi_F)$ the Higgs
potential. Note that when taking $g_U = g_F = g_Y\equiv g_o$, we
arrive at an MSMUM with a single coupling constant $g_o$.

Here we are able to issue some fundamental questions: \\ 1) as
quarks and leptons belong to the Majorana-Weyl representation of 14D
quantum space-time, their motion has been shown to be emergent only
in the 4D space-time. This naturally answers two important issues:
why our living space-time that consists of quarks and leptons is
4-dimensional, and why parity becomes no invariance in our world
that is made of quarks and leptons;\\ 2) fermionic building blocks
in the MSMUM are twice as those in the standard model. The
additional building blocks are regarded as mirroquarks and
mirroleptons, which can be seen more explicitly from the symmetry
decomposition $ SU(16)_W \times SU(16)_E\times U(1) \in SU(32) $. Of
particular, when the mirroparticles form stable states that only
weakly interact with the ordinary matter, the corresponding
mirromatter can become interesting candidate of dark matter observed
in our universe;\\ 3) unlike to the usual grand unification
theories, proton may become rather stable in the MSMUM. This is
because the subgroup $SU(16)_W \in SU(32)$ provides a maximal gauge
symmetry among quarks and leptons\cite{PSS}, so that the gauge
interactions for all quarks and leptons are associated with
different gauge bosons. When $SU(16)_W$ is appropriately broken down
to the symmetries in the standard model without causing, in the mass
eigenstates, a mixing among gauge bosons that can mediate proton
decays, then proton remains stable. Namely the stability of proton
in the MSMUM relies on whether a mixing occurs among relevant gauge
bosons in the mass eigenstates; \\ 4) when the constraints from
proton stability become weak, some symmetry breaking scales can be
low in the MSMUM. It is specially interesting to look for new
particles at TeV scales. But possible intermediate energy scales and
symmetry breaking scenarios should
match to the constraints from the running coupling constants; \\
5) as the Majorana condition in the MSMUM leads to a
self-Hermician Lagrangian, it then implies that
CP symmetry should be broken down spontaneously\cite{SCPV,SCPVW1,SCPVW2}; \\
6) since the symmetry is maximal in the MSMUM with a minimal
parameter, it is expected to be more predictive. Especially, the
minimal scalar field $\Phi_F$ in the Yukawa coupling contains $(10
+ \bar{10} + 126 + \bar{126})$ representations needed for mass
generation and see-saw mechanism, which helps to understand from
vacuum structures of symmetry breaking pattern how
quarks and leptons get masses and mixing, and why neutrinos are so light;\\
7) like grand unification models, symmetry breaking scenarios and
vacuum structures will be the most important issues in the MSMUM.
A simple symmetry breaking scenario can be:  $ SU(32) \to SU(16)_W
\times SU(16)_E \times U(1) $ \\  $SU(16)_W\to SU(8)_L\times
SU(8)_R \to SU(4)_L \times SU(2)_L \times SU(4)_R \times SU(2)_R
\to SU(3)_c \times SU(2)_L \times U(1)_Y\to SU(3)_c \times
U(1)_{em}$. The properties of mirror particles will depend on the
symmetry breaking patterns of $SU(16)_E$. In general, for breaking
a maximal symmetry to a symmetry of real world, it is crucial to
apply for suitable symmetry breaking mechanisms, like the
dynamically spontaneous symmetry breaking realized in
QCD\cite{DSSB}; \\ 8) when extending the principle of
MSMU-hypothesis to the space-time symmetry, we shall arrive at
supersymmetric MSMUM.

   Last but not least, we would like to address that the above
resulting 4-dimensional MSMUM naturally match to the so-called
no-go theorem proved by Coleman and Mendula\cite{NOGO}. The main
assumption made here is that the Majorana fermions belonging to
the spinor representation of 14D quantum space-time are equally
treated on the same footing and directly identified to the basic
building blocks of quarks and leptons. In general, there are many
possibilities for imposing different Majorana spinor structures.
For different spinor structures, one then arrives at different
geometries of high dimensional space-time, which leads to
different physics. What we have demonstrated in this note is that
when choosing the Majorana spinor structure in the 14D space-time
to be directly identified to the quarks and leptons, it then
automatically leads to a unique solution with kinetic term only
appearing in the 4D space-time for each generation quarks and
leptons, the remaining 10D space becomes an internal space without
motions. This may be regarded as an alternative dimension
reduction via choosing the Majorana spinor structure based on the
observations in the real world of elementary particles. In
general, without requiring a specific Majorana spinor structure in
a high dimensional space-time to be assigned to the observed
quarks and leptons, the Majorana spinor fermions can in principle
have motions in the whole high dimensional space-time, namely
there should be no constraints for the dimensions of motion in the
most general case. One may compare such a reduction of dimension
of motion via specifying a spinor structure with the
compactification approach in which the high dimensions are
compacted to lead to a 4D space-time. For different
compactifications, one yields different spinor structures. The
well-known example is the string theory, where the Majorana
spinors belonging the representations in the 10D space-time are
not required to directly relate to the quarks and leptons, so that
they can have motions in the whole 10D space-time. For such a
case, one needs to make an appropriate compactification to yield
an effective 4D space-time theory. There exist in general many
patterns for the compactifications in the string theories.
Obviously, different compactifications lead to different spinor
structures which are corresponding to different physics.

In conclusion, starting from a simple hypothesis, we are led to a
4-dimensional MSMUM with the gauge symmetry $G_{M}^{4D} =
SO(1,3)\times SU(32)\times U(1)_A \times SU(3)_F$.

Acknowledgement: This work was supported in part by the projects
of NSFC and Chinese Academy of Sciences. The author would like to
thank Henry Tye for valuable comments.


\begin{thebibliography}{99}
\bibitem{SM1}  S.L. Glashow, "Partial Symmetries Of Weak Interactions", Nucl. Phys. {\bf 22}, 579 (1961).
\bibitem{SM2} S. Weinberg, "A Model Of Leptons", Phys. Rev. Lett. {\bf 19}, 1264 (1967).
\bibitem{SM3} A. Salam, in
{\it Proceedings of the Eight Nobel Symposium}, edited by N.
Svartholm (Almqvist and Wikell, Stockholm, 1968).
\bibitem{GM} M. Gell-Mann,  "A Schematic model of baryons and mesons", Phys. Lett. 8 214 (1964).
\bibitem{GZ} G. Zweig,  "An SU(3) model for strong interaction symmetry and its breaking", CERN preprint TH401 (1964)
\bibitem{PV} T.D. Lee and C.N. Yang,  "Question Of Parity Conservation In Weak Interactions", Phys. Rev. 104 254 (1956).
\bibitem{YM} C.N. Yang and R.L. Mills,  "Conservation of isotopic spin and isotopic gauge invariance", Phys. Rev. {\bf 96}, 191 (1954)
\bibitem{AF1}D.J. Gross and F. Wilczek,  "Ultraviolet Behavior Of Nonabelian Gauge Theories ", Phys. Rev. Lett., 30  1343
(1973).
\bibitem{AF2} H.D. Polizer, "Reliable Perturbative Results For Strong Interactions?", Phys. Rev. Lett., 30  1346 (1973).
\bibitem{DR} G.'t Hooft and M. Veltman,  "Combinatorics of gauge fields ", Nucl. Phys. {\bf B50}  318 (1972).
\bibitem{RC} H. Georgi, H.R. Quinn and S. Weinberg,  "Hierarchy Of Interactions In Unified Gauge Theories ", Phys. Rev.
Lett. {\bf 33} 451 (1974).
\bibitem{SU422} J. Parti and A. Salam, "Lepton Number As The Fourth Color", Phys.Rev. {\bf D10}  275 (1974).
\bibitem{SU5} H. Georgi and S.L. Glashow, "Unity Of All Elementary Particle Forces", Phys. Rev. Lett. {\bf
32} 438 (1974).
\bibitem{SO10G} H. Georgi, in {\it Particles and Fields}- 1974,
ed. C. Carlson (Amer. Inst. of Physics, New York, 1975).
\bibitem{SO10} H. Fritzsch and P. Minkowski, "Unified Interactions Of Leptons And Hadrons ", Ann. Phys. {\bf 93}, 193 (1975).
\bibitem{SO14} K.C. Chou and Y.L. Wu, "A Unification Model for All Basic Forces", Science in China, Series {\bf A41} 324
(1998).
\bibitem{PSS} J. C. Pati, A. Salam and J.A. Strathdee,  "Can SU(16) Be Ruled Out By Low-Energy Experiments?", Phys. Lett. {\bf B108} 121 (1982).
\bibitem{SCPV} T.D. Lee, "A Theory Of Spontaneous T Violation ", Phys. Rev. D8, 1226 (1973).
\bibitem{SCPVW1} Y.L. Wu and L. Wolfenstein, "Sources of CP violation in the two Higgs doublet model", Phys. Rev. Lett. 73, 1762 (1994); L.
\bibitem{SCPVW2} Wolfenstein and Y.L. Wu,  "CP violation in the decay b ' s gamma in the two Higgs doublet model ", Phys. Rev. Lett., 73, 2809 (1994)£»
\bibitem{DSSB} Y.B. Dai and Y.L. Wu, "Dynamically Spontaneous Symmetry Breaking and Masses of
Lightest Nonet Scalar Mesons as Composite Higgs Bosons", Eur.Phys.J. C39 (2005) S1, hep-ph/0304075.
\bibitem{NOGO} S. Coleman and J. Mandula,  "All Possible Symmetries Of The S Matrix", Phys. Rev. {\bf 159}, 1251 (1967).
\end{thebibliography}
\end{document}